\documentclass[a4paper,11pt]{JHEP}
\usepackage{amssymb}

\newcommand{\be}{\begin{equation}}
\newcommand{\ee}{\end{equation}}
\newcommand{\bea}{\begin{array}}
\newcommand{\ea}{\end{array}}
\newcommand{\beqa}{\begin{eqnarray}}
\newcommand{\eeqa}{\end{eqnarray}}
\newcommand{\bean}{\begin{eqnarray*}}
\newcommand{\eean}{\end{eqnarray*}}

\newcommand{\gad}{\mbox{gad}}

\newcommand{\nn}{\nonumber}

\def\up#1{\leavevmode \raise.16ex\hbox{#1}}

\def\BI{{\rm 1\!l}}

\title{SUSY Anomalies Break $\mathcal{N}=2$ to $\mathcal{N}=1$: The Supersphere and the Fuzzy Supersphere}

\author{Aiyalam P. Balachandran \\
    Physics Department, Syracuse University \\
        Syracuse, NY, 13244-1130, USA \\
    E-mail: \email{bal@phy.syr.edu}}

\author{Aleksandr Pinzul \\
    Physics Department, Syracuse University \\
        Syracuse, NY, 13244-1130, USA \\
    E-mail: \email{apinzul@physics.syr.edu}}

\author{Babar Qureshi \\
    Physics Department, Syracuse University \\
        Syracuse, NY, 13244-1130, USA \\
    E-mail: \email{bqureshi@physics.syr.edu}}

\preprint{SU-4525-812}

\abstract{The $\mathcal{N}=1$ SUSY on $S^2$ and its fuzzy
finite-dimensional matrix version (see
\cite{Landi:1999zz,Hasebe:2004yp} and references therein) are
known. The latter regulates quantum field theories, and seems
suitable for numerical work and capable of higher dimensional
generalizations. In this paper, we study their instanton sectors.
They are SUSY generalizations of $U(1)$ bundles on $S^2$ and their
fuzzy versions, and can be characterized by $k\in\mathbb{Z}$, the
SUSY Chern numbers. In the no-instanton sector ($k=0$),
$\mathcal{N}=2$ SUSY can be chirally realized, the 3 new
$\mathcal{N}=2$ generators anticommuting with the ``Dirac''
operator defining the free action. If $k\neq 0$, the Dirac
operator has zero modes which form an $\mathcal{N}=1$
supermultiplet and an atypical representation of $\mathcal{N}=2$
SUSY. They break the chiral SUSY generators by the Fujikawa
mechanism
\cite{Fujikawa:1979ay,Fujikawa:1980eg,Balachandran:1981cs}. We
have not found this mechanism for SUSY breakdown in the
literature. All these phenomena occur also on the supersphere
SUSY, the graded commutative limit of the fuzzy model. We plan to
discuss that as well in a later work.}


\begin{document}

\section{Overview}

In this section, we give an overview of fuzzy SUSY as full details
can be found elsewhere \cite{inprep}. In later sections, we will
explain all the requisite details to develop instanton theory.

\subsection{The Fuzzy Sphere}

We recall that the fuzzy sphere $S^2_F (n)$ is the $(n+1)\times
(n+1)$ matrix algebra $Mat(n+1)$. It can be realized as linear
operators on $\mathcal{H}^{n+1}$ with the orthonormal basis
vectors \be \label{basis} \frac{(a^\dag_1)^{n_1}}{\sqrt{n_1
!}}\frac{(a^\dag_2)^{n_2}}{\sqrt{n_2 !}}|0\rangle\ ,\ \ n_1 + n_2
= n\ ,\ee where $a_i,\ a^\dag_i$ are bosonic oscillators. The
vectors (\ref{basis}) span a subspace of the Fock space with fixed
particle number $n$: \be N:=\sum_i a^\dag_i a_i\ ,\ \
N|_{\mathcal{H}^{n+1}}=n\ .\ee In this representation, the
elements of $S^2_F (n)$ are the linear operators
$$
\sum_{i,j} c^m_{i,j}(a^\dag_i)^m (a_j)^m\ ,\ \  c^m_{i,j}\in
\mathbb{C}\ ,
$$
restricted to the subspace $\mathcal{H}^{n+1}$.

The group $SU(2)$ acts on $\mathcal{H}^{n+1}$ and hence on $S^2_F
(n)$ by its spin $\frac{n}{2}$ unitary irreducible representation.
The angular momentum generators are \be L_i = a^\dag
\frac{\sigma_i}{2}a\ , \ \ \sigma_i \mbox{ are Pauli matrices.}\ee

\subsection{SUSY}

The $\mathcal{N}=1$ SUSY version of $SU(2)$ is $OSp(2,1)$. It has
the graded Lie algebra $osp(2,1)$. Its generators (basis) can be
written using oscillators if we introduce one additional fermionic
oscillator $b$ and its adjoint $b^\dag$. They commute with $a_i,\
a^\dag_j$. Then the $osp(2,1)$ generators are
\begin{eqnarray}
\Lambda_i & = & a^\dag \frac{\sigma_i}{2}a\ , \ \ \Lambda_4 =
-\frac{1}{2}(a^\dag_1 b + b^\dag a_2)\ ,\\
\Lambda_5 & = & \frac{1}{2}(-a^\dag_2 b + b^\dag a_1)\ ,\ \
\sigma_i \mbox{ = Pauli matrices.} \nonumber
\end{eqnarray}

The $\mathcal{N}=2$ SUSY version of $SU(2)$ is $OSp(2,2)$. It has
the graded Lie algebra $osp(2,2)$. Its basis consists of the
$osp(2,1)$ generators and three additional generators
\begin{eqnarray}
\Lambda_{4'} & \equiv & \Lambda_6 = \frac{1}{2}(a^\dag_1 b -
b^\dag a_2)\ , \ \ \Lambda_{5'} \equiv \Lambda_7 =
\frac{1}{2}(a^\dag_2 b +
b^\dag a_1)\ ,\\
\Lambda_8 & = & a^\dag a + 2 b^\dag b\ .\nonumber
\end{eqnarray}

If $\{\cdot,\cdot\}$ denotes the anticommutator, $osp(2,2)$ has
the defining relations
\begin{eqnarray}
[\Lambda_i , \Lambda_j] & = & i\varepsilon_{ijk}\Lambda_k\ ,\ \
[\Lambda_i , \Lambda_\alpha]=\frac{1}{2}\Lambda_\beta
(\sigma_i)_{\beta\alpha}\ ,\ \ \{\Lambda_\alpha , \Lambda_\beta
\}=\frac{1}{2}(\varepsilon\sigma_i)_{\alpha\beta}\Lambda_i\ ,\\
\left[\Lambda_i , \Lambda_8\right] & = & 0\ ,\ \ [\Lambda_\alpha ,
\Lambda_8]=-\Lambda_{\alpha'}\ ,\ \ \{\Lambda_\alpha ,
\Lambda_{\alpha'}\}=\frac{1}{4}\varepsilon_{\alpha\beta}\Lambda_8\
, \nonumber\\
\{\Lambda_{\alpha'} , \Lambda_{\beta'} \} & = &
-\frac{1}{2}(\varepsilon\sigma_i)_{\alpha\beta}\Lambda_i\ ,\ \
[\Lambda_{\alpha'} , \Lambda_8]=-\Lambda_{\alpha}\ ,\ \
\varepsilon = \left(
\begin{array}{cc}
0 & 1 \\
-1 & 0
\end{array}
\right)\ .\nonumber
\end{eqnarray}
These relations show in particular that the additional three
generators form a triplet under $osp(2,1)$.

Conventional Lie algebras like that of $su(2)$ have a $*$ or an
adjoint operation ${}^\dag$ defined on them. For $\Lambda_i$, it
is just $\Lambda_i^\dag=\Lambda_i$. This follows from the fact
that $a_i^\dag$ is the adjoint of $a_i$. For $osp(2,1)$ and
$osp(2,2)$, ${}^\dag$ is replaced by the grade adjoint ${}^\ddag$.
On the oscillators, ${}^\ddag$ is defined by
$$
a_i^\ddag = a_i^\dag\ ,\ \ (a_i^\dag)^\ddag = (a_i^\dag)^\dag =
a_i\ ,\ \ b^\ddag = b^\dag\ ,\ \ (b^\ddag)^\ddag = -b\ .
$$
Hence ${\ddag}={\dag}$ on bosonic oscillators.

On products of operators, ${}^\ddag$ is defined as follows. We
assign the grade 0 to $a_i,\ a_j^\dag$ and their products and 1 to
$b$ and $b^\dag$. The grades are additive (mod 2). The grade of an
operator $L$ with definite grade is denoted by $|L|$. Then if $L,\
M$ have definite grades, $(LM)^\ddag \equiv (-1)^{|L||M|}M^\ddag
L^\ddag$. Hence $(b^\dag b)^\ddag = b^\dag b$ and \be\label{conj}
\Lambda_i^\ddag=\Lambda_i\ ,\ \
\Lambda_\alpha^\ddag=-\varepsilon_{\alpha\beta}\Lambda_\beta\ ,\ \
\Lambda_{\alpha'}^\ddag=\varepsilon_{\alpha\beta}\Lambda_{\beta'}\
\  \Lambda_{8}^\ddag = \Lambda_{8}\ . \ee

\subsection{Irreducible Representations}

Let $osp(2,0)$ denote $su(2)$, the Lie algebra of $SU(2)$. Its
IRR's are $\Gamma_J^0$, $J\in \mathbb{N}/2$. (Here
$\mathbb{N}=\{0,1,2,...\}$.) $J$ has the meaning of angular
momentum.

The $osp(2,1)$ algebra is of rank 1 just as $osp(2,0)$. We can
take $\Lambda_3$ to be the generator of its Cartan subalgebra.
Since
$$
[\Lambda_3,\Lambda_4]=\frac{1}{2}\Lambda_4\ ,\ \
[\Lambda_3,\Lambda_+ =\Lambda_1 +i\Lambda_2]=\Lambda_+\ ,
$$
$\Lambda_4,\ \Lambda_+$ are its raising operators. They commute:
$$
[\Lambda_4,\Lambda_+]=0\ .
$$
In an IRR, both vanish on the highest weight vector. The
eigenvalue $J\in\mathbb{N}/2$ of $\Lambda_3$ on the highest weight
vector can be used to label its IRR's. They are denoted by
$\Gamma_J^1$ in this paper.

When restricted to its subalgebra $osp(2,0)$, $\Gamma^1_J$ splits
as follows: \be\label{dec21}
\left.\Gamma_J^1\right|_{osp(2,0)}=\Gamma_J^0
\oplus\Gamma_{J-\frac{1}{2}}^0\ ,\ \ J\geq \frac{1}{2}\ . \ee
$\Gamma_0^1$ is the trivial IRR.

The dimension of $\Gamma_J^1$ is $4J+1$.

The graded Lie algebra $osp(2,2)$ is of rank 2. A basis for its
Cartan subalgebra is $\{\Lambda_3\ ,\Lambda_8\}$. Since
$$
[\Lambda_3,\Lambda_4 + \Lambda_{4'}] = \frac{1}{2}(\Lambda_4 +
\Lambda_{4'})\ ,\ \ [\Lambda_8,\Lambda_4 + \Lambda_{4'}] =
\Lambda_4 + \Lambda_{4'}\ ,
$$
$\Lambda_4 + \Lambda_{4'}$ serves as the raising operator for both
$\Lambda_3$ and $\Lambda_8$. We also have that $\Lambda_1
+i\Lambda_2 = \Lambda_+$ is the raising operator for $\Lambda_3$
alone:
$$
[\Lambda_3,\Lambda_+]=\Lambda_+\ ,\ \ [\Lambda_8,\Lambda_+]=0\ .
$$
The raising operators $\Lambda_4 + \Lambda_{4'}$ and $\Lambda_+$
commute:
$$
[\Lambda_4 + \Lambda_{4'},\Lambda_+]=0\ .
$$
Both vanish on the highest weight vector in an IRR while the
eigenvalues $J\in\mathbb{N}/2$ and $k\in\mathbb{Z}$ of $\Lambda_3$
and $\Lambda_8$ on the highest weight vector can be used as labels
of the IRR. They are denoted in this paper by $\Gamma_J^2(k)$.

The $osp(2,2)$ IRR's fall into classes, the \textit{typical} and
\textit{atypical} (or \textit{short}) IRR's. In the typical IRR's,
$|k|\neq 2J$ or $k=J=0$, while in the atypical IRR's, $|k|=2J\neq
0$. The typical IRR with $|k|\neq 2J$ restricted to $osp(2,1)$
splits as follows:
$$
\left.\Gamma_J^2(k)\right|_{osp(2,1)}=\Gamma_J^1
\oplus\Gamma_{J-\frac{1}{2}}^1\ ,\ \ J\geq \frac{1}{2}\ .
$$
$\Gamma_0^2(0)$ is the trivial representation.

The atypical IRR's $\Gamma_J^2(\pm \frac{J}{2})$ ($J\geq 1/2$)
remain irreducible on restriction to $osp(2,1)$:
$$
\left.\Gamma_J^2(\pm {J}/{2})\right|_{osp(2,1)}=\Gamma_J^1\ .
$$
$\Gamma_J^2(\pm {J}/{2})$ can also be abbreviated to
$\Gamma_{J\pm}^2$:
$$
\Gamma_J^2(\pm {J}/{2}) \equiv \Gamma_{J\pm}^2\ ,\ \ J\geq 1/2\ .
$$

$osp(2,2)$ admits the automorphism \be\label{auto} \tau :\
\Lambda_i \rightarrow \Lambda_i\ ,\ \ \Lambda_\alpha \rightarrow
\Lambda_\alpha\ ,\ \ \Lambda_{\alpha'} \rightarrow
-\Lambda_{\alpha'}\ ,\ \ \Lambda_8 \rightarrow -\Lambda_8 \ee
which interchanges $\Gamma_{J}^2(\pm k)$:
$$
\tau :\ \Gamma_{J}^2(k)\rightarrow\Gamma_{J}^2(-k)\ .
$$

\subsection{Casimir Operators}

The $osp(2,0):= su(2)$ Casimir operator $K_0$ is well-known:
$$
K_0 = \Lambda_i^2\ .
$$

The $osp(2,1)$ Casimir operator is
$$
K_1 = \Lambda_i^2 +
\varepsilon_{\alpha\beta}\Lambda_\alpha\Lambda_\beta\ .
$$
We have that
$$
\left. K_1\right|_{\Gamma_J^1}=J(J+\frac{1}{2})\BI\ .
$$

The $osp(2,2)$ quadratic Casimir operator is \be\label{k2} K_2 =
K_1 - \varepsilon_{\alpha\beta}\Lambda_{\alpha'}\Lambda_{\beta'}-
\frac{1}{4}\Lambda_8^2 := K_1 -V_0\ . \ee It has the property
\begin{eqnarray}\label{valk2}
\left. K_2\right|_{\Gamma_J^2(k)} & = & J^2 - \frac{k^2}{4}\ ,
\nonumber\\
\left. K_2\right|_{\Gamma_{J\pm}^2} & = & 0\ .
\end{eqnarray}
As already mentioned, the IRR's $\Gamma_{J\pm}^2$ can be
distinguished by the sign of $\Lambda_8$ on the highest weight
vector.

$osp(2,2)$ also has a cubic Casimir operator
\cite{Frappat:1996pb}, but we will not have occasion to use it.

\subsection{Tensor Products}

The basic Clebsh-Gordan series we need to know is as follows:
$$
\Gamma_J^1\otimes \Gamma_K^1 =
\Gamma_{J+K}^1\oplus\Gamma_{J+K-1/2}^1\oplus\cdots\oplus\Gamma_{|J-K|}^1\
.
$$

\subsection{The Supertrace and the Grade Adjoint}

Because of the decomposition (\ref{dec21}), the vector space
$\mathbb{C}^{4J+1}$ on which $\Gamma_J^1$ acts can be written as
$\mathbb{C}^{2J+1}\oplus\mathbb{C}^{2J}$ where the first term has
angular momentum $J$ and the second term has angular momentum
$J-1/2$. By definition, the first term is the even subspace and
the second term is the odd subspace. The supertrace $str$ of a
matrix
$$
M= \left(%
\begin{array}{cc}
  P_{(2J+1)\times (2J+1)} &\ Q_{(2J+1)\times 2J} \\
  R_{2J\times (2J+1)} & S_{2J\times 2J} \\
\end{array}%
\right)
$$
is accordingly
$$
str M = trP -trS\ .
$$

The grade adjoint $M^\ddag$ can be calculated using the rules of
graded vector spaces \cite{Scheunert:1976wi}. The result is
$$
M^\ddag= \left(%
\begin{array}{cc}
  P^\dag & -R^\dag \\
  Q^\dag & S^\dag \\
\end{array}
\right)
$$
This formula is coherent with (\ref{conj}).

If $Q,R=0$, we say that $M$ is even, while if $P,S=0$, we say that
$M$ is odd. We assign a number $|M|=0,1$ (mod 2) to even and odd
matrices $M$ respectively.

\subsection{The Free Action}

The space with $N=n$ has maximum angular momentum $J=n/2$. It
carries the $osp(2,1)$ IRR $\Gamma_{n/2}^1$ which splits under
$su(2)$ into $\Gamma_{n/2}^0\oplus\Gamma_{(n-1)/2}^0$. It carries
either of the short $osp(2,2)$ IRR's as well.

The dimension of the Hilbert space with $N=n$ is $2n+1$. We denote
it by $\mathcal{H}^{2n+1}$. It is the direct sum
$\mathcal{H}^{n+1}\oplus\mathcal{H}^{n}$ where $\mathcal{H}^{n+1}$
is the even subspace carrying the IRR $\Gamma_{n/2}^0$ and
$\mathcal{H}^{n}$ is the odd subspace carrying the representation
$\Gamma_{(n-1)/2}^0$. A basis for $\mathcal{H}^{2n+1}$ is
\be\label{basis21} \frac{(a^\dag_1)^{n_1}}{\sqrt{n_1
!}}\frac{(a^\dag_2)^{n_2}}{\sqrt{n_2 !}}(b^\dag)^{n_3}|0\rangle\
,\ \ \sum n_i = n\ ,\ \ n_3\in (0,1) \mbox{ , } (b^\dag)^0 :=\BI\
. \ee

The fuzzy SUSY $S_F^{2,2}$ (in the zero instanton sector) is the
matrix algebra $Mat(4J+1)=Mat(2n+1)$. Just as $S_F^2$, it can be
realized using oscillators. In terms of oscillators, a typical
element is
$$
\sum_{i,j} c^m_{i,j}(a^\dag_i)^m (a_j)^m\  + \sum_{i,j}
d^{m-1}_{i,j}(a^\dag_i)^{m-1} (a_j)^{m-1}b^\dag b\ ,\ \
 c^m_{i,j},d^{m-1}_{i,j}\in
\mathbb{C}\ .
$$
It is to be restricted to the space $\mathcal{H}^{2n+1}$.

The left- and right-actions
$$
\Lambda_\rho^L M = \Lambda_\rho M\ ,\ \ \Lambda_\rho^R M =
(-1)^{|\Lambda_\rho||M|}M \Lambda_\rho
$$
of $osp(2,\mathcal{N})$ on $Mat(2n+1)$ give two commuting IRR's of
$osp(2,\mathcal{N})$. Here, $\Lambda_\rho \in osp(2,\mathcal{N})\
,\ \mathcal{N}=1,2,\ M\in Mat(2n+1)$ and both $\Lambda_\rho$ and
$M$ are of definite grade $|\Lambda_\rho|,|M|$ (mod 2).

Combining the left- and right- representations, we get the grade
adjoint representation
$$
\gad :\ \Lambda_\rho \rightarrow \gad\Lambda_\rho= \Lambda_\rho^L
- \Lambda_\rho^R\ ,\ \ \rho\in (i,\alpha,\alpha',8)
$$
of $osp(2,\mathcal{N})$.

With regard to $\gad$, $Mat(4J+1)$ transforms as \be\label{dicomp}
\Gamma_J^1\otimes \Gamma_J^1 =
\Gamma_{2J}^1\oplus\Gamma_{2J-1/2}^1\oplus\Gamma_{2J-1}^1\oplus\cdots\oplus\Gamma_{0}^1\
. \ee

$osp(2,2)$ acts on $Mat(4J+1)$ by $L,R$ and $\gad$ representations
as well. The $L$ and $R$ are the short representations
$\Gamma_{J\pm}^2$ so that under $\gad$, $Mat(4J+1)$ transforms as
$\Gamma_{J+}^2\otimes \Gamma_{J-}^2$. Its reduction can be
inferred from (\ref{dicomp}) once we know that
$\Gamma_j^2(0)|_{osp(2,1)}=\Gamma_j^1\oplus\Gamma_{j-1/2}^1$. We
will see this later. Hence

$$
\Gamma_{J+}^2\otimes \Gamma_{J-}^2 =
\Gamma_{2J}^2(0)\oplus\Gamma_{2J-1}^2(0)\oplus\cdots\oplus\Gamma_{0}^2(0)\
.
$$

The fuzzy field $\Phi$ is an element of fuzzy SUSY. The free
action for $\Phi$ is
$$
S_0 = \frac{f^2}{2}str\ \Phi^\ddag V_0 \Phi\ ,
$$
where $f$ is a real constant and $V_0$ is an $osp(2,1)$-invariant
operator. When restricted to the odd subspace, it should become
the Dirac operator of \cite{Grosse:1994ed,Balachandran:2003ay}.

The limit of this operator for $J=\infty$ was found by Fronsdal
\cite{Fronsdal:1986cd} and later used effectively by Grosse
\textit{et al}. \cite{Grosse:1995pr} For $J=\infty$, it is the
difference $K_1 -K_2$ of the Casimir operators $K_1$ and $K_2$
written as graded differential operators. This operator, for
finite $J$, becomes \be\label{v0} V_0 =
\varepsilon_{\alpha\beta}(\Lambda_{\alpha'})(\Lambda_{\beta'}) +
\frac{1}{4}(\Lambda_{8})^2\ . \ee The simplifications of $S_0$ for
this choice of $V_0$ is given elsewhere \cite{inprep}.

It is evident that $V_0$ is $osp(2,1)$-invariant. But it is less
obvious that $\gad\ \Lambda_{\alpha'}$, $\gad\ \Lambda_{8}$
\textit{anti}-commute with $V_0$: \be\label{anti}
\{\gad\Lambda_{\alpha'},V_0\}=\{\gad\Lambda_{8},V_0\}=0\ . \ee
This means that these generators are realized as \textit{chiral}
symmetries. Of these, $\gad\;\Lambda_{8}$, restricted to the odd
sector, is just standard chirality. Thus, these generators,
associated with $osp(2,2)/osp(2,1)$ are SUSY generalizations of
conventional chirality.

We now show these results.

\section{SUSY Chirality}

Let us first exhibit the highest weight vectors of the $su(2)$
IRR's which occur in $\Gamma_j^2(0)$. Here $j$ is an integer.
Referring to (\ref{dec21}), we have that
$\left.\Gamma_j^1\right|_{su(2)} =
\Gamma_j^0\oplus\Gamma_{j-1/2}^0$ for $j\geq 1$. The highest
weight vector of $\Gamma_j^0$ is $(a_1^\dag a_2)^j$ as it commutes
with $\Lambda_4$ and carries the eigenvalue $j$ of $\gad\
\Lambda_3$. $\gad\ \Lambda_5$ maps it to $-j(a_1^\dag
a_2)^{j-1}\Lambda_4$, the highest weight vector of
$\Gamma_{j-1/2}^0\subset\Gamma_{j}^1$. Thus \be\label{diag}
\begin{array}{cccc}
 \ \ \ \ \ \ \ \  \left.\Gamma_j^1\right|_{su(2)} = & \Gamma_j^0 & \oplus & \!\!\!\!\!\!\!\!\!\!\!\!\!\!\!\!\!\!\Gamma_{j-1/2}^0 \\
\!\!\!\!\!\!\!\!\!\left.%
\begin{array}{c}
  \mbox{Highest}\ \mbox{weight} \\
  \mbox{vectors} \\
\end{array}%
\right\} & (a_1^\dag a_2)^j & \stackrel{\mbox{gad}\Lambda_5
}{\longrightarrow} &\ \ \  -j(a_1^\dag a_2)^{j-1}\Lambda_4\ ,\ \
j\geq 1\ .
  \\
\end{array}
\ee The equation also indicates the operator mapping one highest
weight vector of $su(2)$ to the other.

Next consider $\Gamma_{j-1/2}^1 \supset
\Gamma_{j-1/2}^0\oplus\Gamma_{j-1}^0$ for $j\geq 1$. To
distinguish the $su(2)$ IRR's here from those in $\Gamma_{j}^1$,
we put a prime on them:
$$
\left.\Gamma_{j-1/2}^1\right|_{su(2)}=
\Gamma_{j-1/2}^{0'}\oplus\Gamma_{j-1}^{0'}\ .
$$
The highest weight state of $\Gamma_{j-1/2}^1$, commuting with
$\Lambda_4$ and with eigenvalue $j-1/2$ for $\gad\ \Lambda_3$ is
$-j(a_1^\dag a_2)^{j-1}\Lambda_6$. And $\Lambda_5$ maps it to the
highest weight vector $X_{j-1}$ of $\Gamma_{j-1}^{0'}$. We show
$X_{j-1}$ below. Thus
\begin{eqnarray}\label{diagprime}
\begin{array}{cccc}
  \left.\Gamma_{j-1/2}^1\right|_{su(2)} = & \Gamma_{j-1/2}^{0'} & \oplus & \Gamma_{j-1}^{0'} \\
\!\!\!\!\!\!\!\!\!\!\!\!\!\!\!\left.%
\begin{array}{c}
  \mbox{Highest}\ \mbox{weight} \\
  \mbox{vectors} \\
\end{array}%
\right\} & -j(a_1^\dag a_2)^{j-1}\Lambda_6 &
\stackrel{\mbox{gad}\Lambda_5 }{\longrightarrow} & X_{j-1}\ , \\
\end{array}\\
X_{j-1}= \frac{j-2J-1}{4}(a_1^\dag a_2)^{j-1} +
\frac{1-2j}{4}(a_1^\dag a_2)^{j-1}b^\dag b  \ ,\ \ j\geq 1\
.\nonumber
\end{eqnarray}

In calculating $X_{j-1}$, we use
$$
\Lambda_4\Lambda_6 = -\frac{1}{4}(a_1^\dag a_2)(2b^\dag b - 1)\ ,\
\ a^\dag a + b^\dag b = 2J\ .
$$
Now $\gad\ \Lambda_7$, $\gad\ \Lambda_8$ map the vectors in
(\ref{diag}) to the vectors in (\ref{diagprime}). The full table
is \be
\begin{array}{ccccccl}\label{bigdiag}
    &   & \Gamma_j^1 & \ni & (a_1^\dag a_2)^j & \stackrel{\mbox{gad}\Lambda_5
}{\longrightarrow} &\ \ -j(a_1^\dag a_2)^{j-1}\Lambda_4 \\
    & \nearrow &   &  & \in\Gamma_j^0  &   &\ \ \in\Gamma_{j-1/2}^0 \\
  \Gamma_j^2(0)&  &  &  & \gad\Lambda_7\downarrow & \swarrow \gad\Lambda_8 &\ \ \downarrow\gad\Lambda_7  \\
    & \searrow &   &   &   &   &   \\
    &  & \Gamma_{j-1/2}^1 & \ni & -j(a_1^\dag a_2)^{j-1}\Lambda_6 &
\stackrel{\mbox{gad}\Lambda_5 }{\longrightarrow} &\ \  X_{j-1}\ \ \ ,\ \ \ \ \ \ \ j\geq 1\ . \\
  &  &  &  & \in\Gamma_{j-1/2}^{0'}  &  &\ \   \in\Gamma_{j-1}^{0'}\\
\end{array}
\ee

For $j=0$, we get the trivial IRR of $osp(2,\mathcal{N})$'s.

Eq. (\ref{bigdiag}) shows that $\gad\ \Lambda_{\alpha'}$, $\gad\
\Lambda_{8}$ map the vectors of $\Gamma_j^1$ to those of
$\Gamma_{j-1/2}^1$ ($j\geq 1$) and vice versa. So if $V_0$ has
opposite eigenvalues in the representations in (\ref{bigdiag}),
then we can conclude that\footnote{To show that
$\{\gad\Lambda_{6},V_0\}=0$ we use the fact that $\gad\Lambda_6 =
- [\gad\Lambda_4,\gad\Lambda_8]$. The result follows from the
graded Jacobi identity.}
$$
\{\gad\Lambda_{\alpha'},V_0\}=\{\gad\Lambda_{8},V_0\}=0
$$
identically, since $V_0|_{\Gamma_0^1}=0$. That means that these
operators associated with $osp(2,2)/osp(2,1)$ are
\textit{chirally} realized symmetries.

\section{Eigenvalues of $V_0$}

As $V_0$ is an $osp(2,1)$ scalar, it is enough to compute its
eigenvalue on the highest weight state of $\Gamma^1_j$ to find
$V_0|_{\Gamma_j^1}$.

As $\Lambda_{4'}=\Lambda_6$ commutes with $(a_1^\dag a_2)^j$, we
have that
$$
\varepsilon_{\alpha\beta}\;\gad\Lambda_{\alpha'}\;
\gad\Lambda_{\beta'}\;(a_1^\dag a_2)^j = (\;\gad\Lambda_{4'}\;
\gad\Lambda_{5'}+\gad\Lambda_{5'}\;\gad\Lambda_{4'}\;)\;(a_1^\dag
a_2)^j
$$
where the sign of the second term has been switched as it is zero
anyway. Thus the left-hand side of the previous formula can be
written as
$$
\gad\{\Lambda_{4'},\Lambda_{5'}\}(a_1^\dag a_2)^j =
\frac{1}{2}\gad\Lambda_3 (a_1^\dag a_2)^j = \frac{j}{2}(a_1^\dag
a_2)^j\ .
$$
Also
$$
\gad\Lambda_8(a_1^\dag a_2)^j=0\ .
$$
Hence \be\label{v01} V_0(a_1^\dag a_2)^j=\frac{j}{2}(a_1^\dag
a_2)^j\ . \ee

One quick way to evaluate $V_0|_{\Gamma_{j-1/2}^1}$ is as follows.
Since $K_1|_{\Gamma_{j}^1}=j(j+1/2)$, we have \be
K_2|_{\Gamma_{j}^1}=(K_1-V_0)|_{\Gamma_{j}^1}=j^2\ .\ee But $K_2$
is $osp(2,2)$-invariant. Hence
$$
K_2|_{\Gamma_{j-1/2}^1}=j^2\ .
$$
Since also $K_1|_{\Gamma_{j-1/2}^1}=j(j-1/2)$, we have
$$
V_0|_{\Gamma_{j-1/2}^1}= (K_1 -K_2)|_{\Gamma_{j-1/2}^1} =
-\frac{j}{2}\BI\ .
$$
Thus $V_0$ has opposite eigenvalues on $\Gamma_{j}^1$ and
$\Gamma_{j-1/2}^1$ .

It is important to notice that
$$
K_2 = (2V_0)^2\ .
$$
That is, $2V_0$ is a square root of $K_2$, a bit in the way that
the Dirac operator is the square root of the Laplacian.

\section{Fuzzy SUSY Instantons}

The manifold $S^2$ admits twisted $U(1)$ bundles labelled by a
topological index or Chern number $k\in \mathbb{Z}$. In the
algebraic language, sections of vector bundles associated with
these $U(1)$ bundles are described by elements of projective
modules \cite{Gracia-Bondia:2001tr,Baez:1998he}.

When $S^2$ becomes the graded supersphere $S^{2,2}$, we expect
these modules to persist, and become in some sense supersymmetric
projective modules. That is in fact the case. We shall see that
explicitly after first studying their fuzzy analogues.

The projective modules on $S^2$ and $S^2_F$ are associated with
$SU(2)\simeq S^3$ via Hopf fibration and Lens spaces. In the same
way, the supersymmetric projective modules on $S^{2,2}$ and
$S^{2,2}_F$ get associated with $osp(2,1)$ and $osp(2,2)$.

The fuzzy algebra $S^{2,2}_F$ of previous sections is to be
assigned $k=0$. The elements of this algebra are square matrices
mapping the space with $N=2J$ to the same space $N=2J$. We
emphasize the value of $k$ by writing $S^{2,2}_F(0)$ for
$S^{2,2}_F$. $S^{2,2}_F(0)$ is a bimodule for $osp(2,2)$ as the
latter can act on the left or right of $S^{2,2}_F(0)$ by the IRR's
$\Gamma_{J\pm}^2(0)$.

For $k\neq 0$, $S^{2,2}_F(k)$ is not an algebra. It can be
described using projectors \cite{Baez:1998he,Balachandran:2003ay}
or equally well as maps of the vector space with $N=2J$ to the one
with $N=2J+k$ \cite{Grosse:1995pr}. (We take $J+\frac{k}{2}\geq
0$. If $k<0$, this means $J\geq \frac{|k|}{2}$.) If a basis is
chosen for domain and range of $S^{2,2}_F(k)$, their elements
become rectangular matrices with $2J+k$ rows and $2J$ columns.
$S^{2,2}_F(k)$ as well is a bimodule for $osp(2,2)$. The latter
acts by $\Gamma_{(J+\frac{k}{2})+}^2$ on the left of
$S^{2,2}_F(k)$ and by $\Gamma_{J-}^2$ on the right of
$S^{2,2}_F(k)$.

The invariant associated with $S^{2,2}_F(k)$ is just $k$. The
meaning of $k$ is
$$
k=\mbox{Dimension of range of } S^{2,2}_F(k) - \mbox{Dimension of
domain of } S^{2,2}_F(k)\ .
$$

Scalar fields $\Phi$ are now elements of $S^{2,2}_F(k)$ while
$V_0$ is replaced by a new operator $V_k$ which incorporates the
appropriate connection and ``topological'' data. We now argue,
using index theory and other considerations, that the
$osp(2,1)$-invariant $V_k$ is fixed by the requirement
$$
V_k^2 = K_2
$$
where $K_2$ is the Casimir invariant for
$\Gamma_{(J+\frac{k}{2})+}^2\otimes\Gamma_{J-}^2$.

\section{Fuzzy SUSY Zero Modes and their Index Theory}

We begin by analyzing the $osp(2,1)$ and $osp(2,2)$ representation
content of $S^{2,2}_F(k)$.

As regards the $\gad$ representation of $osp(2,1)$, it transforms
according to
$$
\Gamma^1_{J+\frac{k}{2}}\otimes\Gamma^1_{J} =
\left(\Gamma^1_{2J+\frac{k}{2}}\oplus\Gamma^1_{2J+\frac{k}{2}-\frac{1}{2}}\right)\oplus
\left(\Gamma^1_{2J+\frac{k}{2}-1}\oplus\Gamma^1_{2J+\frac{k}{2}-\frac{3}{2}}\right)
\oplus\cdots\oplus
\left(\Gamma^1_{\frac{|k|}{2}+1}\oplus\Gamma^1_{\frac{|k|}{2}+\frac{1}{2}}\right)
\oplus\Gamma^1_{\frac{|k|}{2}}\ .
$$

The analogue of (\ref{bigdiag}) is:
\begin{eqnarray}\label{bigdiagk}
2J+\frac{k}{2}\geq j \geq \frac{|k|}{2} + 1\ , \nn \\
\nn \\
\begin{array}{ccccccc}
                &          & \Gamma_j^1       & \ni & \Gamma_j^0       & \oplus & \Gamma_{j-1/2}^0 \\
                & \nearrow &                  &     &                  &        &                  \\
  \Gamma_j^2(k) &          &                  &     &                  &        &                  \\
                & \searrow &                  &     &                  &        &                  \\
                &          & \Gamma_{j-1/2}^1 & \ni & \Gamma_{j-1/2}^0 & \oplus & \Gamma_{j-1}^0\ \ \ \ . \\
\end{array}
\end{eqnarray}
Here $|k|\geq 1$. For $j=\frac{|k|}{2}$, we get the atypical
representation of $osp(2,2)$:
$$
\Gamma_{\frac{|k|}{2}}^2(k)\rightarrow \Gamma_{\frac{|k|}{2}}^1 =
\Gamma_{\frac{|k|}{2}}^0\oplus \Gamma_{\frac{|k|}{2}-1}^0\ .
$$

All this becomes explicit during the following calculation of the
eigenvalues of $K_2$.

\subsection{Spectrum of $K_2$}

For $k>0$, the highest weight vector with angular momentum
$$
j=m+\frac{|k|}{2}\ ,\ \ m=0,1,...
$$
is
$$
(a_1^\dag)^{|k|}(a_1^\dag a_2)^m \ .
$$
Since
$$
\gad\Lambda_8 (a_1^\dag)^{|k|}(a_1^\dag a_2)^m =
|k|(a_1^\dag)^{|k|}(a_1^\dag a_2)^m \ ,
$$
it is the highest weight vector of $\Gamma_j^2(|k|)$. Thus
$\Gamma_j^2(|k|)$ occurs in the reduction of the $osp(2,2)$ action
on $S_F^{2,2}(|k|)$.

General theory \cite{Frappat:1996pb} tells us the branching rules
of $\Gamma_j^2(|k|)$ as in (\ref{bigdiagk}). This equation is thus
established for $k>0$.

We can check as before that
$$
\varepsilon_{\alpha\beta}\;\gad\Lambda_{\alpha'}\;\gad\Lambda_{\beta'}\;(a_1^\dag)^{|k|}(a_1^\dag
a_2)^m =
\frac{1}{2}\left(m+\frac{|k|}{2}\right)(a_1^\dag)^{|k|}(a_1^\dag
a_2)^m
$$
while
$$
\frac{1}{4}(\gad\Lambda_8)^2(a_1^\dag)^{|k|}(a_1^\dag a_2)^m =
\frac{k^2}{4}
$$
and
$$
K_1|_{\Gamma_j^1}=j(j+1)\BI\ .
$$
We thus have \cite{Frappat:1996pb}
$$
K_2|_{\Gamma_j^2(|k|)} = \left(j^2 - \frac{k^2}{4}\right)\BI\ .
$$

For $k<0$,
$$
(a_2)^{|k|}(a_1^\dag a_2)^m
$$
is the highest weight vector for angular momentum
$$
j= m + \frac{|k|}{2}\ .
$$
Since
$$
\gad\Lambda_8 (a_2)^{|k|}(a_1^\dag a_2)^m =
-|k|(a_2)^{|k|}(a_1^\dag a_2)^m \ ,
$$
it is the highest weight vector of $\Gamma_j^2(-|k|)$. Hence
$\Gamma_j^2(-|k|)$ occurs in the reduction of the $osp(2,2)$
action on $S_F^{2,2}(-|k|)$. We thus establish (\ref{bigdiagk})
for $k<0$ as well.

The eigenvalues of $V_k$, when restricted to $\Gamma_j^1$ and
$\Gamma_{j-1/2}^1$ and for $j\geq \frac{|k|}{2}+1$, are
$\pm\sqrt{j^2-\frac{k^2}{4}}$. These eigenvalues are not zero.
Hence the $osp(2,2)$ operators which intertwine these
representations, mapping vectors of one representation to the
other, \textit{anticommute} with $V_k$: they are \textit{chiral}
symmetries for these representations. For $j=\frac{|k|}{2}$, $V_k$
vanishes while the representation space carries the atypical
representation $\Gamma_{\frac{|k|}{2}}^2 (k)$ of $osp(2,2)$. Hence
we can say that the above chiral operators all anticommute with
$\left. V_k\right|_{J=0}$. Hence these operators anticommute with
$V_k$ (for any $j$, on all vectors of $S_F^{2.2}(k)$) just as
standard chirality anticommutes with the massless Dirac operator.

For $k=0$, these operators were $\Lambda_{\alpha'}$, $\Lambda_8$.
But they change with $k$. They can be worked out. They do not
occur in subsequent discussion and hence we do not show them here.

We now establish that $V_k$ is the correct choice of the action
for the fuzzy SUSY action $S_k$: \be\label{vk} S_k =
\mbox{const}\;str\;\Phi^\ddag\;V_k\;\Phi \ . \ee This formula is
valid also for $k=0$ as we saw earlier. We here focus on $k\neq
0$.

The Dirac operators $D$ for fuzzy spheres of instanton number $k$
are known \cite{Balachandran:2003ay}. We first show that $V_k$
coincides with this operator on the Dirac sector.

It is enough to focus on typical $osp(2,2)$ IRR's since both the
Dirac operator and $V_k$ vanish on the grade-odd sector of
$\Gamma^1_{\frac{|k|}{2}}$. Thus consider $\Gamma^2_j (k)$ for
$j\geq \frac{1}{2}|k|+1$. Angular momentum $J$ in the Dirac sector
of $\Gamma^2_j (k)$ is $j-1/2$. Hence
$$
\left. V^2_k \right|_{\Gamma^2_j (k)\ Dirac\ sector} = \left(
J-\frac{|k|-1}{2} \right) \left( J+\frac{|k|+1}{2} \right)\BI \ .
$$
Substituting $J= n + \frac{|k|-1}{2}$ and identifying $|k| = 2T$,
we get the answer of \cite{Balachandran:2003ay}:
$$
\left. V^2_k \right|_{\Gamma^2_j (k)\ Dirac\ sector} = n(n+2T)\BI
\ .
$$
Hence $\left. V^2_k \right|_{\Gamma^2_j (k)\ Dirac\ sector}$ is
the correct Dirac operator .

This result and the $osp(2,1)$-invariance of $V_k$ are compelling
reasons to identify it as the SUSY generalization of the Dirac and
Laplacian operators for $k\neq 0$.

\subsection{Index Theory and Zero Modes}

There is also further evidence supporting the correctness of
$V_k$: It gives the SUSY generalization of index theory.

Thus one knows that 1) the Dirac operator has $|k|$ zero modes for
instanton number $k$ on $S^2$ and on the fuzzy sphere $S^2_F (k)$,
\cite{Balachandran:2003ay,Balachandran:1999hx} and 2) they are
left- (right-) chiral if $k>0$ ($k<0$), 3) charge conjugation
interchanges these chiralities.

More precisely if $n_{L,R}$ are the number of left- and
right-chiral zero modes,
$$
n_L - n_R = k\ .
$$
This number is ``topologically stable''. The meaning of this
statement in the fuzzy case can be found in
\cite{Balachandran:2003ay}.

If the Dirac operator is $SU(2)$-invariant, these zero modes
organize themselves into $SU(2)$ multiplets with angular momentum
$\frac{|k|}{2}$ \cite{Balachandran:2003ay,Balachandran:1999hx}.

Now $V_k$ has zero modes which form the atypical multiplet
$\Gamma^2_{\frac{|k|}{2}}(k)$ of $osp(2,2)$. The number of zero
modes is $2|k|+1$. Of these, $|k|$ correspond to the grade odd
sector and can be identified with the zero modes of $S^2$ and
$S^2_F (k)$ Dirac operators. The remaining grade even ($|k|+1$)
zero modes are their SUSY-partners.

The zero modes transform by inequivalent IRR's of $osp(2,2)$ for
the two signs of $k$. These two atypical $osp(2,2)$
representations are SUSY generalizations of left- and right-
chiralities.

Identifying charge conjugation with the automorphism (\ref{auto}),
we see that it exchanges these two IRR's just as it exchanges
chiralities in the Dirac sector.

\section{Final Remarks}

In this paper, we have extended the work of \cite{inprep} on the
fuzzy SUSY model on $S^2$ to the instanton sector. A SUSY
generalization of index theory and zero modes of the Dirac
operator has also been established.

Following \cite{inprep}, we can try introducing interactions
involving just $\Phi$. For $k\neq 0$, $\Phi$ can be thought of as
a rectangular matrix. So $\Phi^\ddag\Phi$ and $\Phi\Phi^\ddag$ are
square matrices of different sizes acting on $osp(2,2)$
representations with $N=n$ and $N=n+k$. A typical interaction may
then be \be\label{interact} \lambda str (\Phi^\ddag\Phi)^2 \ee
where $str$ is over the space with $N=n$, the domain of
$\Phi^\ddag\Phi$. (But note that (\ref{interact}) and the use of
$str$ in interactions require further study.)

Fuzzy SUSY gauge theories remain to be formulated. The
investigation of the graded commutative limit $n\rightarrow\infty$
with $k$ fixed has also not been done for $k\neq 0$.

Numerical simulations on fuzzy SUSY models are being initiated.

\bigskip
\bigskip

{\parindent 0cm{\bf Acknowledgement}}

We are part of a collaboration on fuzzy physics and have benefited
by inputs from its members. We are particulary grateful to Marco
Panero for carefully reading the paper and suggesting several
corrections. This work was supported in part by the DOE grant
DE-FG02-85ER40231 and by NSF under contract number INT9908763.


\bigskip
\bigskip


\begin{thebibliography}{99}

\bibitem{Landi:1999zz}
  G.~Landi,
  Differ.\ Geom.\ Appl.\  {\bf 14}, 95 (2001)
  [arXiv:math-ph/9907020].

\bibitem{Hasebe:2004yp}
  K.~Hasebe and Y.~Kimura,
  Nucl.\ Phys.\ B {\bf 709}, 94 (2005)
  [arXiv:hep-th/0409230] and references therein.

\bibitem{Fujikawa:1979ay}
  K.~Fujikawa,
  Phys.\ Rev.\ Lett.\  {\bf 42}, 1195 (1979).

\bibitem{Fujikawa:1980eg}
  K.~Fujikawa,
  Phys.\ Rev.\ D {\bf 21}, 2848 (1980)
  [Erratum-ibid.\ D {\bf 22}, 1499 (1980)].

\bibitem{Balachandran:1981cs}
  A.~P.~Balachandran, G.~Marmo, V.~P.~Nair and C.~G.~Trahern,
  Phys.\ Rev.\ D {\bf 25}, 2713 (1982).

\bibitem{inprep}
  A.~P.~Balachandran, B.~Dolan, D.~O'Connor, M.~Panero and
  S.~K\"{u}rk\c{c}\"{u}o\v{g}lu, in preparation.

\bibitem{Frappat:1996pb}
  L.~Frappat, P.~Sorba and A.~Sciarrino,
  ``Dictionary on Lie superalgebras,''
  arXiv:hep-th/9607161.

\bibitem{Scheunert:1976wi}
  M.~Scheunert, W.~Nahm and V.~Rittenberg,
  J.\ Math.\ Phys.\  {\bf 18}, 146 (1977).

\bibitem{Fronsdal:1986cd}
  C.~Fronsdal, in
  ``Essays On Supersymmetry'', ed. by C.~Fronsdal, D.~Reidel
  Publishing Company, Dordrecht, 1986.

\bibitem{Grosse:1995pr}
  H.~Grosse, C.~Klimcik and P.~Presnajder,
  Commun.\ Math.\ Phys.\  {\bf 185}, 155 (1997)
  [arXiv:hep-th/9507074].

\bibitem{Grosse:1994ed}
  H.~Grosse and P.~Presnajder,
  Lett.\ Math.\ Phys.\  {\bf 33}, 171 (1995).

\bibitem{Balachandran:2003ay}
  A.~P.~Balachandran and G.~Immirzi,
  Phys.\ Rev.\ D {\bf 68}, 065023 (2003)
  [arXiv:hep-th/0301242].

\bibitem{Balachandran:1999hx}
  A.~P.~Balachandran and S.~Vaidya,
  Int.\ J.\ Mod.\ Phys.\ A {\bf 16}, 17 (2001)
  [arXiv:hep-th/9910129].

\bibitem{Gracia-Bondia:2001tr}
  J.~M.~Gracia-Bondia, J.~C.~Varilly and H.~Figueroa,
  ``Elements of noncommutative geometry'',  Boston, USA: Birkhaeuser,
  2001.

\bibitem{Baez:1998he}
  S.~Baez, A.~P.~Balachandran, B.~Ydri and S.~Vaidya,
  Commun.\ Math.\ Phys.\  {\bf 208}, 787 (2000)
  [arXiv:hep-th/9811169].

\end{thebibliography}
\end{document}